\documentclass[prd,nofootinbib,preprintnumbers]{revtex4}
\usepackage[utf8]{inputenc}
\usepackage{amsfonts,amsmath,amssymb}
\usepackage{graphicx}
\usepackage{color}
\usepackage[plainpages=false, colorlinks=true, anchorcolor=blue, linkcolor=blue, citecolor=blue, bookmarks=false]{hyperref}
\usepackage{natbib}
\usepackage{enumitem}
\usepackage{subcaption}
\newcommand{\rthis}[1]{\textcolor{black}{#1}}
\begin{document}
\newcommand{\apjl}{Astrophys. J. Lett.}
\newcommand{\apjs}{Astrophys. J. Suppl. Ser.}
\newcommand{\aap}{Astron. \& Astrophys.}
\newcommand{\aj}{Astron. J.}
\newcommand{\araa}{Ann. Rev. Astron. Astrophys. } 
\newcommand{\mnras}{Mon. Not. R. Astron. Soc.}
\newcommand{\apss} {Astrophys. and Space Science}
\newcommand{\jcap}{JCAP}
\newcommand{\pasj}{PASJ}
\newcommand{\pasa}{Pub. Astro. Soc. Aust.}
\newcommand{\physrep}{Physics Reports}
\title{Search for Annual  Modulation in Combined ANAIS 112 and COSINE 100 Data using Bayesian Model Comparison}

\author{Om \surname{Godse}$^1$}\altaffiliation{Email: me23btech11045@iith.ac.in}

\author{Shantanu  \surname{Desai}$^2$ }  
\altaffiliation{Email: shntn05@gmail.com}

\affiliation{$^{1}$ Department of Mechanical and Aerospace  Engineering, Indian Institute of Technology Hyderabad, Telangana 502285, India}

\affiliation{$^{2}$Department of Physics, Indian Institute of Technology Hyderabad, Telangana 502285, India}

\begin{abstract}
We perform a Bayesian model comparison test between a  sinusoidal  modulation model  and a constant value model using about three years of combined COSINE-100 and ANAIS-112 data. 
We use both uniform priors and normal priors (based on DAMA best-fit values) for the angular frequency and phase of the cosine signal. We find natural log of Bayes factor for the cosine model compared to the constant value model to be less than 1.15 for the data in both 1-6 keV and 2-6 keV energy intervals.   This shows that there is no evidence for cosine signal from dark matter interactions  in the combined ANAIS-112/COSINE-100 data.  Our analysis codes have also been made publicly available.
\end{abstract}

\maketitle

\section{Introduction}
Even though we know for more than 90 years that most of the mass of our galaxy is made up of  cold dark matter, the identity of the dark matter candidate(s) is still a mystery~\cite{Bertone}. One of the most widely studied dark matter candidate is the Weakly interacting massive particle (WIMP), since such a candidate with weak-scale interactions has a relic abundance which matches that of the observed dark matter density~\cite{Weinberg}. A large number of experiments (refered to as direct detection experiments) have been looking for signatures of dark matter scattering off atomic nuclei for close to four decades~\cite{Goodman}.

Among the plethora of direct detection experiments, only the DAMA/LIBRA  experiment has reported  positive evidence for dark matter interactions. The characteristic signature seen in this experiment is an annual modulation in the residual rates, which has the right characteristics expected from dark matter interactions~\cite{Freese88}. This signal was first announced in 1998~\cite{DAMA98} and the statistic significance of the annual modulation using the latest data with a total exposure of 2.46 ton-year covering more than 14 years is about 12.9$\sigma$~\cite{DAMA18}. However, this signal could not be independently confirmed by other direct detection experiments~\cite{Arcadi} as well as indirect detection experiments~\cite{Desai04}. To resolve this long-standing DAMA anomaly, two experiments namely COSINE-100 and ANAIS-112 have been designed using the same target material as DAMA, viz. NaI. 

Recently, both  COSINE-100 and ANAIS-112   presented results on the search for an  annual modulated based on  joint analysis of their data covering three years~\cite{CosAn} (C25, hereafter). The best-fit values for the annual modulation amplitudes found using least-squares and Markov Chain Monte Carlo based regression analysis were equal to  ($-0.0002 \pm 0.0026$) cpd/kg/keV and 
($0.0021 \pm  0.0028$) cpd/kg/keV  in 1-6 keV and 2-6 keV intervals, respectively. Therefore, these results do now show any evidence for an annual modulation to support the DAMA claim.
In this work, we complement the analysis in ~\cite{CosAn}, by carrying out a Bayesian model comparison test between the hypothesis that the combined COSINE-100/ANAIS-112 data is consistent with annual modulation due to a sinusoidal signal (expected from dark matter interactions), compared to the hypothesis that the data only contain a constant value due to the  background. This work is a follow-up to our previous works on Bayesian comparison for annual modulation in multiple direct dark matter detection experiments such as DAMA~\cite{Krishak,Nikita}, COSINE-100~\cite{Krishak2}, ANAIS-112~\cite{Krishakanais}. Other applications of Bayesian model comparison in analysis of DAMA/LIBRA data can also be found in ~\cite{Messina}.

This  manuscript is structured  as follows.
A brief summary of the combined COSINE-100/ANAIS-112 results discussed in C25 can be found in Sect.~\ref{sec:ANAIS}. Our analysis and results of the same data is described in Sect.~\ref{sec:analysis}.  We conclude in Sect.~\ref{sec:conclusions}. 

\section{Summary of C25}
\label{sec:ANAIS}
We now recap the  results from the analysis in C25, where  more details can be found.  COSINE-100~\cite{Adhikari17} and ANAIS-112~\cite{Anaisdet} are two direct dark matter experiments which use Na(Tl)  as the target  and have been designed to test  the long-standing  DAMA annual modulation claim~\cite{DAMA18}.   The COSINE-100 experiment  is located in the Yangyang underground laboraory in Korea and consists of five NaI(Tl) detectors with a total mass of 61 kg. COSINE-100 has been taking data since October 2016.  Similarly, the ANAIS-112 experiment is located in the Canfranc Underground Laboratory in Spain and consists of nine  NaI(Tl) crystals of mass 12.5 kg each, with a total mass of 112.5 kg. ANAIS-112 has been taking data since August 3, 2017. More details of both these experiments along with background radioactive rates can be found in ~\cite{Adhikari17} and ~\cite{Anaisdet}, for COSINE-100 and ANAIS-112, respectively.

The ANAIS-112 and COSINE-100 collaborations have  published results from about three  years of data and could not support the DAMA/LIBRA annual modulation claim~\cite{Adhikari21,Anaislatest}.
Since both these experiments have similar sensitivity, a combined analysis of both these datasets from the first three years of data  was done in C25. Only the single-hit events were used for analysis. Each dataset was background subtracted to obtain the residuals, which were then combined to do a search for annual modulation.  A separate search was done in both 1-6 keV and 2-6 keV energy intervals. Details of the background models used for both COSINE-100 and ANAIS-12 can be found in C25 and references therein, and will be skipped here for brevity.

The data was sub-divided into 15-day bins for both the experiments. The total exposure analyzed was 485 kg-year.
The background subtracted residuals were fit to the following function:
\begin{equation}
R(t) = S_m \cos (\omega (t-t_0)),
\label{eq:res}
\end{equation}
where $t_0$ is the phase (kept fixed at June 2nd), $\omega=2\pi/T $ for $T=1$ year, and $S_m$ is the modulation amplitude. The best-fit values of $S_m$ were obtained using both frequentist and MCMC analysis.  The best-fit values were equal to  ($-0.0002 \pm 0.0026$) cpd/kg/keV and  ($0.0021 \pm  0.0028$) cpd/kg/keV  in 1-6 keV and 2-6 keV intervals. Therefore, the modulation amplitudes are consistent with no annual modulation to within $1\sigma$. Finally, a similar analysis using  a simple combination of six years of data from COSINE-100~\cite{Cosine6} and ANAIS-112~\cite{Anais6} with a total exposure of 1 ton-year was also carried out in C25. These results confirm the previous results of no modulation, with modulation amplitudes having values of ($0.0005 \pm 0.0019$) cpd/kg/keV and ($0.0027 \pm 0.0021$) cpd/kg/keV in 1-6 keV and 2-6 keV, respectively. 

\section{Analysis and Results}
\label{sec:analysis}
\subsection{Brief Primer on Bayesian Model Comparison}
\label{sec:mc}
We provide a brief prelude to Bayesian model comparison.
and refer the reader to more details in  Refs.~\cite{Trotta,Sanjib,Weller,Krishakeotwash}.

To compare the efficacy of a  ($M_2$) against another model ($M_1$), we define  the Bayes factor ($B_{21}$) given by:
\begin{equation}
B_{21}=    \frac{\int P(D|M_2, \theta_2)P(\theta_2|M_2) \, d\theta_2}{\int P(D|M_1, \theta_1)P(\theta_1|M_1) \, d\theta_1} ,  \label{eq:BF}
\end{equation}
where $P(D|M_2,\theta_2)$ is the likelihood for the model $M_2$ given the data $D$, and $P(\theta_2|M_2)$  is  the prior for  the parameter vector $\theta_2$ of the model $M_2$.   The denominator refers to the same, but for model $M_1$.
 If $B_{21}$ is greater than 1.0, then  model $M_2$ is preferred over $M_1$ and vice-versa. The significance can be evaluated  using Jeffreys' scale~\cite{Weller}.   
 For our analysis, $M_2$ refers to the hypothesis that the residual data can be described using Eq.~\ref{eq:res}, whereas $M_1$ refers to the hypothesis that the residuals can be described by a constant value.

\subsection{Results}
The first step in the calculation of Bayesian  evidence involves assuming a likelihood. For our analysis, we use a Gaussian likelihood assuming Gaussian residuals, which is given by:
 \begin{equation}
     P(D|M)=\prod_{i=1}^N \frac{1}{\sigma_i\sqrt{2\pi}} \exp \left\{-\frac{(\eta_i-f(x,\theta))^2}{2\sigma_i^2}\right\},
  \end{equation}
  where  $\eta_i$ denotes  the observed residuals binned in 15-day intervals; $f(x,\theta)$ is the model function used to fit the data; and $\sigma_i$ denotes  the observed error in $\eta_i$.  For   model $M_2$, $f(x,\theta)$  is given by Eq~\ref{eq:res}.~\footnote{\rthis{Note that in our previous works~\cite{Krishakanais}, we had used  $\omega (t+t_0)$ instead of $\omega (t-t_0)$. Here, we use Eq.~\ref{eq:res} as our signal model to be consistent with C25}.}  For the model  $M_1$, $f(x,\theta)$ is equal to a constant value $A$. The total duration of the dataset is equal to 1381 days  and 1376 days for 1-6 keV and 2-6 keV, respectively.
  The next step involves choosing priors for the free parameters. In C25, only the amplitude was  kept as a free parameter, while all the parameters fixed to that expected from a dark matter signal. For this analysis, we use multiple sets of priors for the amplitude, period, and phase similar to ~\cite{Krishakanais}. To evaluate Bayesian evidence, we use the nested sampler implemented in {\tt Dynesty}~\cite{Speagle}. \rthis{We ran {\tt Dynesty} using dynamic nested sampling with initial number of live points equal to 1024 and $\Delta \ln Z$ tolerance of 0.1, where $Z$ is the Bayesian evidence.} For the amplitude, \rthis{we use two sets of uniform priors for the cosine model. For the first set, we use  uniform priors} between the maximum and minimum value of the observed residuals in 1-6 keV and 2-6 keV. \rthis{We then also considered uniform priors on the cosine amplitude between zero and maximum (absolute) value of the  observed residuals in 1-6 and 2-6 keV. However, in both cases for the constant rate model, the lower bound on the prior corresponds to the minimum value of the observed residuals.}
  For the angular frequency and phase, we use both uniform and normal priors. The uniform priors on the phase are the same  as  in ~\cite{Krishakanais} and given by $t_0 \in [0, 365]$ days. For the uniform prior on $\omega$, we used a lower and upper bound based on the maximum duration of the dataset ($\sim$ 1380 days) and bin size (15 days), respectively.  This corresponds to $\omega \in \mathcal{U}$ (0.00455,0.418) rad/day. 
  The normal prior on $\omega$ was based on the best-fit value for period obtained from the latest DAMA/LIBRA analysis, viz.
  $T=0.999 \pm 0.001 $ \rthis{year}~\cite{DAMA18}, which corresponds to $\omega \in  \mathcal{N} (0.0172,1.36  \times 10^{-5}$) rad/day. Similarly for the phase $t_0$, we used a Gaussian prior based on the  DAMA/LIBRA best-fit value of $t_0 = 145 \pm 5$ days.  A summary of all the four sets of priors for the cosine model can be found in Table~\ref{tab:priortable1}. 
  
\begin{figure*}
\includegraphics[width =0.7\textwidth]{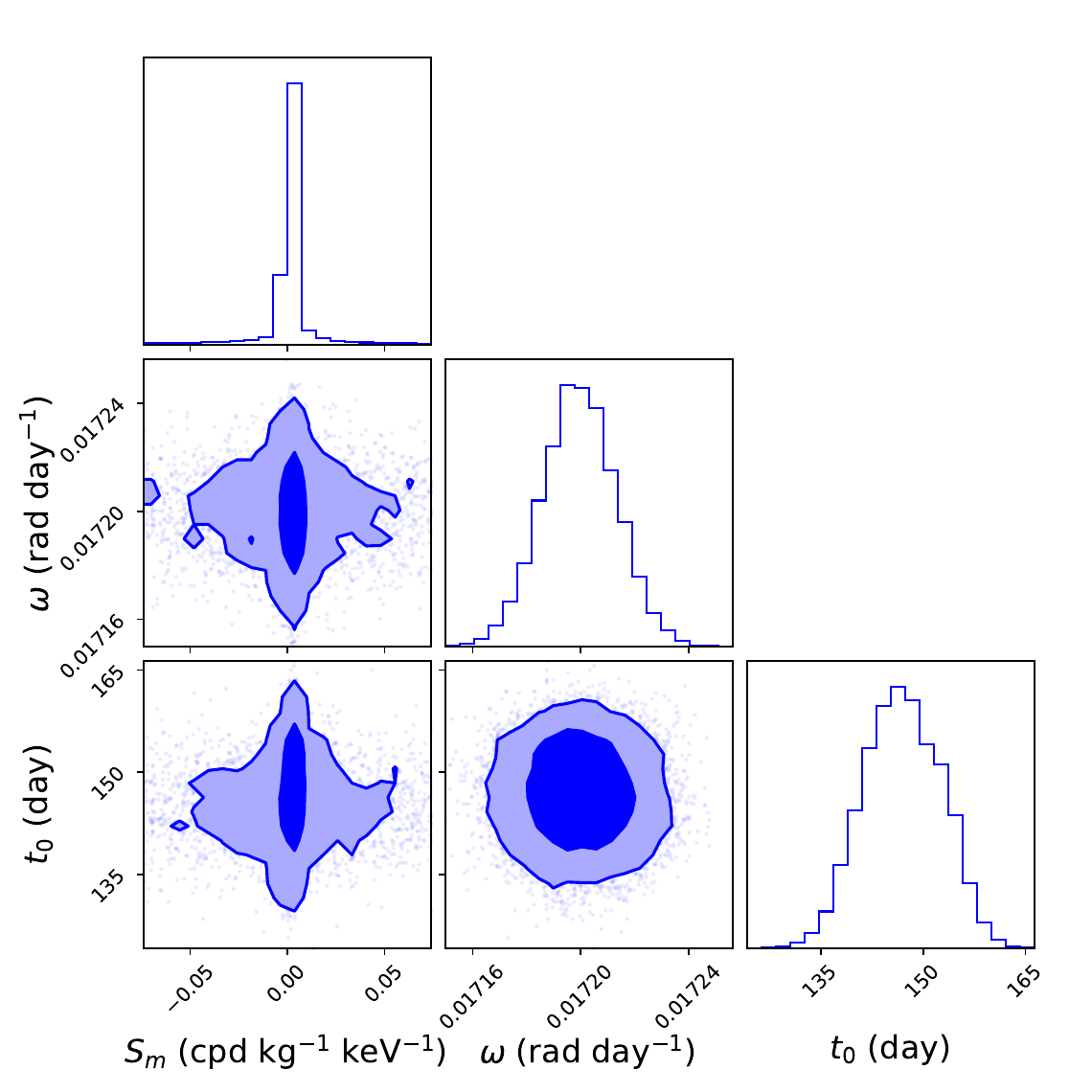}
\caption{ \rthis{Marginalized posterior contours for $S_m$, $\omega$, and $t_0$ for the cosine model (Eq.~\ref{eq:res}) at 68\% and  95\% credible intervals for the residual rates in 2-6 keV energy range using the priors  $\mathcal{U}$ (-0.0789,0.0789) cpd/kg/keV, $\mathcal{N}$ (0.0172,$1.36 \times 10^{-5}$) (rad/day), and $\mathcal{N}$ (145,5) (days) for $S_m$, $\omega$, and $t_0$, respectively.}}

\label{fig:corner}
\end{figure*}

  We then calculate $\ln (B_{21})$ (natural log of Bayes factor) for all the four sets of priors for both 1-6 keV and 2-6 keV energy intervals. These values can be found in last two columns of Table~\ref{tab:priortable1}. We find that in all four cases, $\ln (B_{21}<1.15 (B_{21} \leq \sqrt{10})$, which corresponds to ``barely worth mentioning'' according to Jeffreys scale~\cite{Weller}. Therefore, we conclude that Bayesian model comparison does not provide any evidence to support the hypothesis that the combined residuals from ANAIS-112 and COSINE-100 prefer  annual modulation compared to a constant value. This implies that the results from Bayesian model comparison are consistent with the analysis in C25. \rthis{We also show the marginalized posteriors for each of the three parameters for the cosine model using one of the prior choices for the data in 2-6 keV energy range in Fig.~\ref{fig:corner}.}

\begin{table*}[h]
\begin{tabular}{|ccccc|}
\hline \hline
    Prior on A & Prior on \boldmath$\omega$ & \boldmath $t_0$ Prior & $\ln{B_{21}}$ & $\ln{B_{21}}$
\\
 (cpd/kg/keV) & (radian/day) & (days) & (1 - 6 keV) & (2 - 6 keV) \\ \hline  
 $\mathcal{U}$ (-max($|f_i|$),max($|f_i|$))  & $\mathcal{U}$ (0.00455,0.418)  &$\mathcal{U}$ (0,365) & 1.15 $\pm$ 0.06 & 1.12 $\pm$ 0.06 \\ 
   $\mathcal{U}$ (-max($|f_i|$),max($|f_i|$))  & $\mathcal{U}$ (0.00455,0.418)  &$\mathcal{N}$ (145,5) & 1.14 $\pm$ 0.07 & 1.14 $\pm$ 0.06 \\ 
 $\mathcal{U}$ (-max($|f_i|$),max($|f_i|$))  & $\mathcal{N}$ (0.0172,$1.36 \times 10^{-5}$)  &$\mathcal{U}$ (0,365) & -0.02 $\pm$ 0.05 & 0.73 $\pm$ 0.05 \\ 
 $\mathcal{U}$ (-max($|f_i|$),max($|f_i|$))  & $\mathcal{N}$ (0.0172,$1.36 \times 10^{-5}$)  &$\mathcal{N}$ (145,5)& -0.13 $\pm$ 0.05 & 0.54 $\pm$ 0.05 \\ 
\hline
$\mathcal{U}$ (0,max($|f_i|$))  & $\mathcal{U}$ (0.00455,0.418)  &$\mathcal{U}$ (0,365) & 1.34 $\pm$ 0.07 & 1.15 $\pm$ 0.06 \\ 
   $\mathcal{U}$ (0,max($|f_i|$))  & $\mathcal{U}$ (0.00455,0.418)  &$\mathcal{N}$ (145,5) & 1.36  $\pm$ 0.07 & 1.08 $\pm$ 0.06
 \\ 
 $\mathcal{U}$ (0,max($|f_i|$))  & $\mathcal{N}$ (0.0172,$1.36 \times 10^{-5}$)  &$\mathcal{U}$ (0,365) & 0.04 $\pm$ 0.05 & 0.80 $\pm$ 0.05 \\ 
 $\mathcal{U}$ (0,max($|f_i|$))  & $\mathcal{N}$ (0.0172,$1.36 \times 10^{-5}$)  &$\mathcal{N}$ (145,5)& -0.05 $\pm$ 0.05 & 0.81 $\pm$ 0.05 \\  \hline
\end{tabular}
\caption{Results of Bayesian model comparison for the combined ANAIS-112 and COSINE-100 data for a cosine signal compared to a constant term. The second to fourth columns show the priors used for the cosine model in Eq.~\ref{eq:res}. \rthis{For the null hypothesis, the  prior for the constant term is same as that for $A$ for the first four rows, even though the lower limit on the prior for the cosine model is equal to 0 (last four rows).}  The uniform prior on $\omega$ corresponds to period between 15 and 1380 days. For the prior on $A$, max($|f_i|$) is  equal to 0.0789 for 1-6 keV range and 0.0739 for 2-6 keV range. The last two columns show the Bayes factor  assuming the constant term is the null hypothesis. Based on Jeffreys scale, the significance amounts to ``barely worth mentioning''. }
\label{tab:priortable1}
\end{table*}

\section{Conclusions}
\label{sec:conclusions}

Recently, a joint analysis of  two dark matter detection experiments, namely  COSINE-100 and ANAIS-112,  both using the same target material as the DAMA/LIBRA experiment was published in C25. The analysis is based on 
 the first three years of data, corresponding to a   total exposure of 485 kg.year.  This analysis used Bayesian regression and found that the joint analysis rules out the DAMA annual modulation claim by 4.7$\sigma$ and 3.5$\sigma$ in 1-6 and 2-6 keV energy intervals, respectively.

As a follow-up to  our previous work on Bayesian model comparison  with DAMA, COSINE-100, ANAIS-112  data~\cite{Krishak,Krishak2}, we carried out an independent model comparison analysis for  annual modulation using the same combined data. We calculated the Bayes factors for a  cosine model (expected from dark matter interactions) compared to a constant model for four sets of priors. A tabular summary of our results can be found in Table ~\ref{tab:priortable1}.

We found  that $\ln (B_{21})< 1.15$ for the cosine model compared to a constant model implying extremely marginal or no evidence for an annual modulation over the constant value.  Therefore, Bayesian model comparison  implies that there is no evidence for annual cosine modulation in the combined ANAIS-112/COSINE-100 data from the first three years and agrees with the results in C25.

To improve transparency in data analysis, we have made  our analysis codes public, which  can be found at \url{https://github.com/OmGodse/EPProject1}.

\bibliography{main}

@ARTICLE{Freese88,
       author = {{Freese}, Katherine and {Frieman}, Joshua and {Gould}, Andrew},
        title = "{Signal modulation in cold-dark-matter detection}",
      journal = {\prd},
         year = 1988,
        month = jun,
       volume = {37},
       number = {12},
        pages = {3388-3405},
          doi = {10.1103/PhysRevD.37.3388},
       adsurl = {https://ui.adsabs.harvard.edu/abs/1988PhRvD..37.3388F}
}

@ARTICLE{Goodman,
       author = {{Goodman}, Mark W. and {Witten}, Edward},
        title = "{Detectability of certain dark-matter candidates}",
      journal = {\prd},
         year = 1985,
        month = jun,
       volume = {31},
       number = {12},
        pages = {3059-3063},
          doi = {10.1103/PhysRevD.31.3059},
       adsurl = {https://ui.adsabs.harvard.edu/abs/1985PhRvD..31.3059G}
}

@ARTICLE{Bertone,
       author = {{Bertone}, Gianfranco and {Hooper}, Dan},
        title = "{History of dark matter}",
      journal = {Reviews of Modern Physics},
         year = 2018,
        month = oct,
       volume = {90},
       number = {4},
          eid = {045002},
        pages = {045002},
          doi = {10.1103/RevModPhys.90.045002},
archivePrefix = {arXiv},
       eprint = {1605.04909},
 primaryClass = {astro-ph.CO},
       adsurl = {https://ui.adsabs.harvard.edu/abs/2018RvMP...90d5002B}
}

@ARTICLE{Cosine6,
       author = {{Carlin}, N. and {Cho}, J.~Y. and {Choi}, J.~J. and {Choi}, S. and {Ezeribe}, A.~C. and {Franca}, L.~E. and {Ha}, C. and {Hahn}, I.~S. and {Hollick}, S.~J. and {Jeon}, E.~J. and {Joo}, H.~W. and {Kang}, W.~G. and {Kauer}, M. and {Kim}, B.~H. and {Kim}, H.~J. and {Kim}, J. and {Kim}, K.~W. and {Kim}, S.~H. and {Kim}, S.~K. and {Kim}, W.~K. and {Kim}, Y.~D. and {Kim}, Y.~H. and {Ko}, Y.~J. and {Lee}, D.~H. and {Lee}, E.~K. and {Lee}, H. and {Lee}, H.~S. and {Lee}, H.~Y. and {Lee}, I.~S. and {Lee}, J. and {Lee}, J.~Y. and {Lee}, M.~H. and {Lee}, S.~H. and {Lee}, S.~M. and {Lee}, Y.~J. and {Leonard}, D.~S. and {Luan}, N.~T. and {Machado}, V.~H.~A. and {Manzato}, B.~B. and {Maruyama}, R.~H. and {Neal}, R.~J. and {Olsen}, S.~L. and {Park}, B.~J. and {Park}, H.~K. and {Park}, H.~S. and {Park}, J.~C. and {Park}, K.~S. and {Park}, S.~D. and {Pitta}, R.~L.~C. and {Prihtiadi}, H. and {Ra}, S.~J. and {Rott}, C. and {Shin}, K.~A. and {Cavalcante}, D.~F.~F.~S. and {Son}, M.~K. and {Spooner}, N.~J.~C. and {Truc}, L.~T. and {Yang}, L. and {Yu}, G.~H.},
        title = "{COSINE-100 Full Dataset Challenges the Annual Modulation Signal of DAMA/LIBRA}",
      journal = {arXiv e-prints},
         year = 2024,
        month = sep,
          eid = {arXiv:2409.13226},
        pages = {arXiv:2409.13226},
          doi = {10.48550/arXiv.2409.13226},
archivePrefix = {arXiv},
       eprint = {2409.13226},
 primaryClass = {hep-ex},
       adsurl = {https://ui.adsabs.harvard.edu/abs/2024arXiv240913226C}
}

@ARTICLE{Anais6,
       author = {{Amar{\'e}}, Julio and {Apilluelo}, Jaime and {Cebri{\'a}n}, Susana and {Cintas}, David and {Coarasa}, Iv{\'a}n and {Garc{\'\i}a}, Eduardo and {Mart{\'\i}nez}, Mar{\'\i}a and {Ortigoza}, Ysrael and {de Sol{\'o}rzano}, Alfonso Ortiz and {Pardo}, Tamara and {Puimed{\'o}n}, Jorge and {Sarsa}, Mar{\'\i}a Luisa and {Seoane}, Carmen},
        title = "{Towards a Robust Model-Independent Test of the DAMA/LIBRA Dark Matter Signal: ANAIS-112 Results with Six Years of Data}",
      journal = {\prl},
         year = 2025,
        month = aug,
       volume = {135},
       number = {5},
          eid = {051001},
        pages = {051001},
          doi = {10.1103/ntnl-zrn9},
archivePrefix = {arXiv},
       eprint = {2502.01542},
 primaryClass = {astro-ph.IM},
       adsurl = {https://ui.adsabs.harvard.edu/abs/2025PhRvL.135e1001A}
}

@ARTICLE{CosAn,
       author = {{Carlin}, N. and {Cho}, J.~Y. and {Choi}, J.~J. and {Choi}, S. and {Ezeribe}, A.~C. and {Fran{\c{c}}a}, L.~E. and {Ha}, C. and {Hahn}, I.~S. and {Hollick}, S.~J. and {Hong}, S.~B. and {Jeon}, E.~J. and {Joo}, H.~W. and {Kang}, W.~G. and {Kauer}, M. and {Kim}, B.~H. and {Kim}, H.~J. and {Kim}, J. and {Kim}, K.~W. and {Kim}, S.~H. and {Kim}, S.~K. and {Kim}, W.~K. and {Kim}, Y.~D. and {Kim}, Y.~H. and {Ko}, Y.~J. and {Lee}, D.~H. and {Lee}, E.~K. and {Lee}, H. and {Lee}, H.~S. and {Lee}, H.~Y. and {Lee}, I.~S. and {Lee}, J. and {Lee}, J.~Y. and {Lee}, M.~H. and {Lee}, S.~H. and {Lee}, S.~M. and {Lee}, Y.~J. and {Leonard}, D.~S. and {Luan}, N.~T. and {Machado}, V.~H.~A. and {Manzato}, B.~B. and {Maruyama}, R.~H. and {Neal}, R.~J. and {Olsen}, S.~L. and {Park}, H.~K. and {Park}, H.~S. and {Park}, J.~C. and {Park}, J.~S. and {Park}, K.~S. and {Park}, K. and {Park}, S.~D. and {Pitta}, R.~L.~C. and {Prihtiadi}, H. and {Ra}, S.~J. and {Rott}, C. and {Shin}, K.~A. and {Cavalcante}, D.~F.~F.~S. and {Son}, M.~K. and {Spooner}, N.~J.~C. and {Truc}, L.~T. and {Yang}, L. and {Yu}, G.~H. and {Cosine-100 Collaboration)} and {Amar{\'e}}, J. and {Apilluelo}, J. and {Cebri{\'a}n}, S. and {Cintas}, D. and {Coarasa}, I. and {Garc{\'\i}a}, E. and {Mart{\'\i}nez}, M. and {Ortigoza}, Y. and {Ortiz de Sol{\'o}rzano}, A. and {Pardo}, T. and {Puimed{\'o}n}, J. and {Sarsa}, M.~L. and {Seoane}, C. and {(Anais-112 Collaboration}},
        title = "{Combined Annual Modulation Dark Matter Search with COSINE-100 and ANAIS-112}",
      journal = {\prl},
         year = 2025,
        month = sep,
       volume = {135},
       number = {12},
          eid = {121002},
        pages = {121002},
          doi = {10.1103/9j7w-qp1c},
archivePrefix = {arXiv},
       eprint = {2503.19559},
 primaryClass = {astro-ph.IM},
       adsurl = {https://ui.adsabs.harvard.edu/abs/2025PhRvL.135l1002C}
}

@ARTICLE{Sanjib,
       author = {{Sharma}, Sanjib},
        title = "{Markov Chain Monte Carlo Methods for Bayesian Data Analysis in Astronomy}",
      journal = {\araa},
         year = 2017,
        month = aug,
       volume = {55},
       number = {1},
        pages = {213-259},
          doi = {10.1146/annurev-astro-082214-122339},
archivePrefix = {arXiv},
       eprint = {1706.01629},
 primaryClass = {astro-ph.IM},
       adsurl = {https://ui.adsabs.harvard.edu/abs/2017ARA&A..55..213S}
}

@ARTICLE{Speagle,
       author = {{Speagle}, Joshua S.},
        title = "{dynesty: A Dynamic Nested Sampling Package for Estimating Bayesian Posteriors and Evidences}",
      journal = {\mnras},
         year = 2020,
        month = feb,
          doi = {10.1093/mnras/staa278},
archivePrefix = {arXiv},
       eprint = {1904.02180},
 primaryClass = {astro-ph.IM},
       adsurl = {https://ui.adsabs.harvard.edu/abs/2020MNRAS.tmp..280S}
}

@ARTICLE{Adhikari17,
       author = {{Adhikari}, G. and {Adhikari}, P. and {de Souza}, E. Barbosa and {Carlin}, N. and {Choi}, S. and {Choi}, W.~Q. and {Djamal}, M. and {Ezeribe}, A.~C. and {Ha}, C. and {Hahn}, I.~S. and {Hubbard}, A.~J.~F. and {Jeon}, E.~J. and {Jo}, J.~H. and {Joo}, H.~W. and {Kang}, W.~G. and {Kang}, W. and {Kauer}, M. and {Kim}, B.~H. and {Kim}, H. and {Kim}, H.~J. and {Kim}, K.~W. and {Kim}, M.~C. and {Kim}, N.~Y. and {Kim}, S.~K. and {Kim}, Y.~D. and {Kim}, Y.~H. and {Kudryavtsev}, V.~A. and {Lee}, H.~S. and {Lee}, J. and {Lee}, J.~Y. and {Lee}, M.~H. and {Leonard}, D.~S. and {Lim}, K.~E. and {Lynch}, W.~A. and {Maruyama}, R.~H. and {Mouton}, F. and {Olsen}, S.~L. and {Park}, H.~K. and {Park}, H.~S. and {Park}, J.~S. and {Park}, K.~S. and {Pettus}, W. and {Pierpoint}, Z.~P. and {Prihtiadi}, H. and {Ra}, S. and {Rogers}, F.~R. and {Rott}, C. and {Scarff}, A. and {Spooner}, N.~J.~C. and {Thompson}, W.~G. and {Yang}, L. and {Yong}, S.~H.},
        title = "{Initial performance of the COSINE-100 experiment}",
      journal = {European Physical Journal C},
         year = 2018,
        month = feb,
       volume = {78},
       number = {2},
          eid = {107},
        pages = {107},
          doi = {10.1140/epjc/s10052-018-5590-x},
archivePrefix = {arXiv},
       eprint = {1710.05299},
 primaryClass = {physics.ins-det},
       adsurl = {https://ui.adsabs.harvard.edu/abs/2018EPJC...78..107A}
}

@ARTICLE{Weller,
       author = {{Kerscher}, Martin and {Weller}, Jochen},
        title = "{On model selection in cosmology}",
      journal = {SciPost Physics Lecture Notes},
         year = 2019,
        month = jun,
       volume = {9},
          doi = {10.21468/SciPostPhysLectNotes.9},
archivePrefix = {arXiv},
       eprint = {1901.07726},
 primaryClass = {astro-ph.CO},
       adsurl = {https://ui.adsabs.harvard.edu/abs/2019ScPPL...9.....K}
}

@ARTICLE{Dama18,
       author = {{Bernabei}, Rita and {Belli}, Pierluigi and {Bussolotti}, Andrea and
         {Cappella}, Fabio and {Caracciolo}, Vincenzo and {Cerulli}, Riccardo and
         {Dai}, Chang-Jiang and {d'Angelo}, Annelisa and {Di Marco}, Alessandro and
         {He}, Hui-Lin and {Incicchitti}, Antonella and {Ma}, Xin-Hua and
         {Mattei}, Angelo and {Merlo}, Vittorio and {Montecchia}, Francesco and
         {Sheng}, Xiang-Dong and {Ye}, Zi-Piao},
        title = "{First Model Independent Results from DAMA/LIBRA-Phase2}",
      journal = {Nuclear Physics and Atomic Energy},
         year = "2018",
        month = "Dec",
       volume = {19},
       number = {4},
        pages = {307},
          doi = {10.15407/jnpae2018.04.307},
archivePrefix = {arXiv},
       eprint = {1805.10486},
 primaryClass = {hep-ex},
       
}

@ARTICLE{Krishak2,
       author = {{Krishak}, Aditi and {Desai}, Shantanu},
        title = "{An independent assessment of significance of annual modulation in COSINE-100 data}",
      journal = {The Open Journal of Astrophysics},
         year = "2019",
        month = "Dec",
       volume = {2},
        pages = {E12},
          doi = {10.21105/astro.1907.07199},
archivePrefix = {arXiv},
       eprint = {1907.07199},
 primaryClass = {astro-ph.CO},
       adsurl = {https://ui.adsabs.harvard.edu/abs/2019OJAp....2E..12K}
}

@ARTICLE{Anaisdet,
       author = {{Amar{\'e}}, J. and {Cebri{\'a}n}, S. and {Coarasa}, I. and {Cuesta}, C. and {Garc{\'\i}a}, E. and {Mart{\'\i}nez}, M. and {Oliv{\'a}n}, M.~A. and {Ortigoza}, Y. and {de Sol{\'o}rzano}, A. Ortiz and {Puimed{\'o}n}, J. and {Salinas}, A. and {Sarsa}, M.~L. and {Villar}, P. and {Villar}, J.~A.},
        title = "{Performance of ANAIS-112 experiment after the first year of data taking}",
      journal = {European Physical Journal C},
         year = 2019,
        month = mar,
       volume = {79},
       number = {3},
          eid = {228},
        pages = {228},
          doi = {10.1140/epjc/s10052-019-6697-4},
archivePrefix = {arXiv},
       eprint = {1812.01472},
 primaryClass = {astro-ph.IM},
       adsurl = {https://ui.adsabs.harvard.edu/abs/2019EPJC...79..228A}
}

@ARTICLE{Anaislatest,
       author = {{Coarasa}, Iv{\'a}n and {Amar{\'e}}, Julio and {Apilluelo}, Jaime and {Cebri{\'a}n}, Susana and {Cintas}, David and {Garc{\'\i}a}, Eduardo and {Mart{\'\i}nez}, Mar{\'\i}a and {Oliv{\'a}n}, Miguel {\'A}ngel and {Ortigoza}, Ysrael and {Ortiz de Sol{\'o}rzano}, Alfonso and {Pardo}, Tamara and {Puimed{\'o}n}, Jorge and {Salinas}, Ana and {Sarsa}, Mar{\'\i}a Luisa and {Villar}, Patricia},
        title = "{ANAIS-112 three years data: a sensitive model independent negative test of the DAMA/LIBRA dark matter signal}",
      journal = {Communications Physics},
         year = 2024,
        month = oct,
       volume = {7},
       number = {1},
          eid = {345},
        pages = {345},
          doi = {10.1038/s42005-024-01827-y},
archivePrefix = {arXiv},
       eprint = {2404.17348},
 primaryClass = {astro-ph.IM},
       adsurl = {https://ui.adsabs.harvard.edu/abs/2024CmPhy...7..345C}
}

@ARTICLE{Adhikari21,
       author = {{Adhikari}, G. and {Barbosa de Souza}, E. and {Carlin}, N. and {Choi}, J.~J. and {Choi}, S. and {Ezeribe}, A.~C. and {Fran{\c{c}}a}, L.~E. and {Ha}, C. and {Hahn}, I.~S. and {Hollick}, S.~J. and {Jeon}, E.~J. and {Jo}, J.~H. and {Joo}, H.~W. and {Kang}, W.~G. and {Kauer}, M. and {Kim}, H. and {Kim}, H.~J. and {Kim}, J. and {Kim}, K.~W. and {Kim}, S.~H. and {Kim}, S.~K. and {Kim}, W.~K. and {Kim}, Y.~D. and {Kim}, Y.~H. and {Ko}, Y.~J. and {Kwon}, H.~J. and {Lee}, D.~H. and {Lee}, E.~K. and {Lee}, H. and {Lee}, H.~S. and {Lee}, H.~Y. and {Lee}, I.~S. and {Lee}, J. and {Lee}, J.~Y. and {Lee}, M.~H. and {Lee}, S.~H. and {Lee}, S.~M. and {Leonard}, D.~S. and {Manzato}, B.~B. and {Maruyama}, R.~H. and {Neal}, R.~J. and {Park}, B.~J. and {Park}, H.~K. and {Park}, H.~S. and {Park}, K.~S. and {Park}, S.~D. and {Pitta}, R.~L.~C. and {Prihtiadi}, H. and {Ra}, S.~J. and {Rott}, C. and {Shin}, K.~A. and {Scarff}, A. and {Spooner}, N.~J.~C. and {Thompson}, W.~G. and {Yang}, L. and {Yu}, G.~H. and {Cosine-100 Collaboration}},
        title = "{Three-year annual modulation search with COSINE-100}",
      journal = {\prd},
         year = 2022,
        month = sep,
       volume = {106},
       number = {5},
          eid = {052005},
        pages = {052005},
          doi = {10.1103/PhysRevD.106.052005},
archivePrefix = {arXiv},
       eprint = {2111.08863},
 primaryClass = {hep-ex},
       adsurl = {https://ui.adsabs.harvard.edu/abs/2022PhRvD.106e2005A}
}

@ARTICLE{Krishak,
       author = {{Krishak}, Aditi and {Dantuluri}, Aisha and {Desai}, Shantanu},
        title = "{Robust model comparison tests of DAMA/LIBRA annual modulation}",
      journal = {\jcap},
         year = 2020,
        month = feb,
       volume = {2020},
       number = {2},
          eid = {007},
        pages = {007},
          doi = {10.1088/1475-7516/2020/02/007},
archivePrefix = {arXiv},
       eprint = {1906.05726},
 primaryClass = {astro-ph.CO},
       adsurl = {https://ui.adsabs.harvard.edu/abs/2020JCAP...02..007K}
}

@ARTICLE{Nikita,
       author = {{Bhagvati}, Srinikitha and {Desai}, Shantanu},
        title = "{Bayesian analysis of time dependence of DAMA annual modulation amplitude}",
      journal = {\jcap},
         year = 2021,
        month = sep,
       volume = {2021},
       number = {9},
          eid = {022},
        pages = {022},
          doi = {10.1088/1475-7516/2021/09/022},
archivePrefix = {arXiv},
       eprint = {2106.06724},
 primaryClass = {astro-ph.CO},
       adsurl = {https://ui.adsabs.harvard.edu/abs/2021JCAP...09..022B}
}

@ARTICLE{Krishakanais,
       author = {{Krishak}, Aditi and {Desai}, Shantanu},
        title = "{An independent search for annual modulation and its significance in ANAIS-112 data}",
      journal = {Progress of Theoretical and Experimental Physics},
         year = 2020,
        month = aug,
       volume = {2020},
       number = {9},
          eid = {093F01},
        pages = {093F01},
          doi = {10.1093/ptep/ptaa102},
archivePrefix = {arXiv},
       eprint = {1910.05096},
 primaryClass = {astro-ph.CO},
       adsurl = {https://ui.adsabs.harvard.edu/abs/2020PTEP.2020i3F01K}
}

@ARTICLE{Krishakeotwash,
       author = {{Krishak}, Aditi and {Desai}, Shantanu},
        title = "{Model comparison tests of modified gravity from the E{\"o}t-Wash experiment}",
      journal = {\jcap},
         year = 2020,
        month = jul,
       volume = {2020},
       number = {7},
          eid = {006},
        pages = {006},
          doi = {10.1088/1475-7516/2020/07/006},
archivePrefix = {arXiv},
       eprint = {2003.10127},
 primaryClass = {gr-qc},
       adsurl = {https://ui.adsabs.harvard.edu/abs/2020JCAP...07..006K}
}

@ARTICLE{Messina,
       author = {{Messina}, Andrea and {Nardecchia}, Marco and {Piacentini}, Stefano},
        title = "{Annual modulations from secular variations: not relaxing DAMA?}",
      journal = {\jcap},
         year = 2020,
        month = apr,
       volume = {2020},
       number = {4},
          eid = {037},
        pages = {037},
          doi = {10.1088/1475-7516/2020/04/037},
archivePrefix = {arXiv},
       eprint = {2003.03340},
 primaryClass = {hep-ex},
       adsurl = {https://ui.adsabs.harvard.edu/abs/2020JCAP...04..037M}
}

@ARTICLE{Trotta,
       author = {{Trotta}, Roberto},
        title = "{Bayesian Methods in Cosmology}",
      journal = {arXiv e-prints},
         year = "2017",
        month = "Jan",
          eid = {arXiv:1701.01467},
        pages = {arXiv:1701.01467},
archivePrefix = {arXiv},
       eprint = {1701.01467},
 primaryClass = {astro-ph.CO},
       adsurl = {https://ui.adsabs.harvard.edu/abs/2017arXiv170101467T}
}

@ARTICLE{Weinberg,
   author = {{Lee}, B.~W. and {Weinberg}, S.},
    title = "{Cosmological lower bound on heavy-neutrino masses}",
  journal = {Physical Review Letters},
     year = 1977,
    month = jul,
   volume = 39,
    pages = {165-168},
      doi = {10.1103/PhysRevLett.39.165},
   adsurl = {http://adsabs.harvard.edu/abs/1977PhRvL..39..165L}
}

@article{Desai04,
      author         = "Desai, S. and others",
      title          = "{Search for dark matter WIMPs using upward through-going
                        muons in Super-Kamiokande}",
      collaboration  = "Super-Kamiokande",
      journal        = "Phys. Rev.",
      volume         = "D70",
      year           = "2004",
      pages          = "083523",
      doi            = "10.1103/PhysRevD.70.083523",
      eprint         = "hep-ex/0404025",
      archivePrefix  = "arXiv",
      primaryClass   = "hep-ex",
      SLACcitation   = "%%CITATION = HEP-EX/0404025;%%"
}
\end{document}